\documentclass[prd,superscriptaddress,a4paper,showpacs,showkeys,10pt,nofootinbib]{revtex4}
\usepackage{graphicx}
\usepackage{graphics}
\usepackage{epsfig}
\usepackage{dcolumn}
\usepackage{bm}
\usepackage{multirow}
\usepackage{tabularx}
\usepackage{hyperref}
\usepackage{commath}
\usepackage{epstopdf}
\usepackage[T1]{fontenc}
\usepackage{geometry}
\usepackage{color}

\def\lsim{\raise0.3ex\hbox{$<$\kern-0.75em\raise-1.1ex\hbox{$\sim$}}}

\def\gsim{\raise0.3ex\hbox{$>$\kern-0.75em\raise-1.1ex\hbox{$\sim$}}}

\def\pom{{I\!\!P}}

\newcommand{\be}{\begin{equation}}

\newcommand{\ee}{\end{equation}}

\def\beq{\begin{equation}}

\def\eeq{\end{equation}}

\def\beqa{\begin{eqnarray}}

\def\eeqa{\end{eqnarray}}

\newcommand{\rd}{\mbox{\boldmath $\Delta$}}

\newcommand{\ba}{\begin{eqnarray}}

\newcommand{\ea}{\end{eqnarray}}

\newcommand{\rr}{\mbox{\boldmath $r$}}

\newcommand{\rb}{\mbox{\boldmath $b$}}

\def\gappeq{\mathrel{\rlap {\raise.5ex\hbox{$>$}}

{\lower.5ex\hbox{$\sim$}}}}

\def\lappeq{\mathrel{\rlap{\raise.5ex\hbox{$<$}}

{\lower.5ex\hbox{$\sim$}}}}

\def\Toprel#1\over#2{\mathrel{\mathop{#2}\limits^{#1}}}

\def\pom{{I\!\!P}}

\begin{document}

\title{Coherent and incoherent $J/\Psi$ photoproduction in $PbPb$ collisions \\ at the LHC, HE -- LHC and FCC}
\author{Victor P. Gon\c{c}alves}
\email[]{barros@ufpel.edu.br}
\affiliation{High and Medium Energy Group, Instituto de F\'{\i}sica e Matem\'atica,  Universidade Federal de Pelotas (UFPel)\\
Caixa Postal 354,  96010-900, Pelotas, RS, Brazil. \\
}

\author{Daniel E. Martins}
\email[]{dan.ernani@gmail.com}
\affiliation{Instituto de F\'isica, Universidade Federal do Rio de Janeiro (UFRJ), 
Caixa Postal 68528, CEP 21941-972, Rio de Janeiro, RJ, Brazil}

\author{Celso R. Sena}
\email[]{celsorodriguessena@gmail.com}
\affiliation{High and Medium Energy Group, Instituto de F\'{\i}sica e Matem\'atica,  Universidade Federal de Pelotas (UFPel)\\
Caixa Postal 354,  96010-900, Pelotas, RS, Brazil. \\
}

\begin{abstract}
The coherent and incoherent $J/\Psi$ photoproduction in $PbPb$ collisions are investigated considering the possible states of nucleon configurations in the nuclear wave function and taking into account of the non - linear corrections to the QCD dynamics.
The cross sections and the rapidity and transverse momentum distributions are estimated for the energies of the next run of the LHC,   High -- Energy LHC and Future Circular Collider. Our results indicate that a future experimental analysis of these processes will be  useful to discriminate between different approaches for the QCD dynamics as well to improve our description of the gluon saturation effects.
\end{abstract}



\maketitle

\vspace{1cm}

\section{Introduction}

Heavy ion collisions are an important laboratory that provides distinct alternatives to probe fundamental aspects of the strong interactions theory -- the Quantum Chromodynamics (QCD). For central and semi -- central collisions, where the impact parameter $b$ of the collision is smaller than the sum of the nuclear radius and the strong interactions dominate,  we can probe the creation  of the Quark - Gluon Plasma (QGP) and constrain its properties \cite{Busza:2018rrf}. In contrast,  ultraperipheral collisions (UPHICs), which are defined as  collisions at large impact 
parameters $b > 2 R$ where  the long range photon -- induced interactions become dominant, can be used  to contrain the QCD dynamics at high energies (small -- $x$) \cite{upc}.  One has that the description of the  initial conditions for the collective behaviour of the medium produced in central and semi - central  heavy ion collisions are  determined by the momentum and spatial distributions of gluons in the nuclei, which are expected to be sensitive to the presence of non -- linear effects in the QCD dynamics \cite{hdqcd}. Therefore, there is a strict connection between the physics probed in central, semi - central and ultraperipheral collisions, which can be explored in order to improve our understanding of QCD at large energies and high densities \cite{Schlichting:2019abc}.

\begin{figure}[t]
\begin{tabular}{cc}

 {\includegraphics[width=0.5\textwidth]{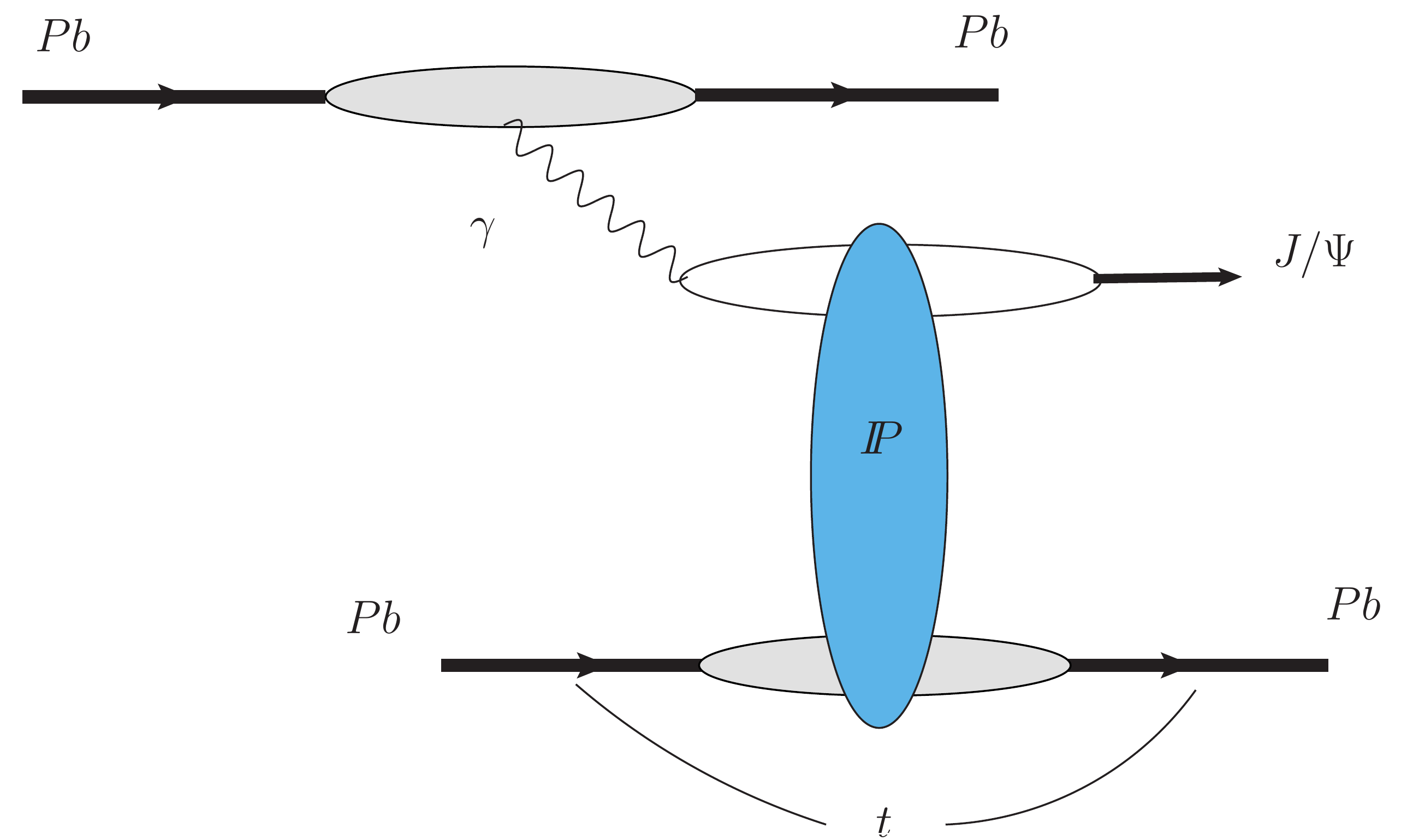}} & 
{\includegraphics[width=0.5\textwidth]{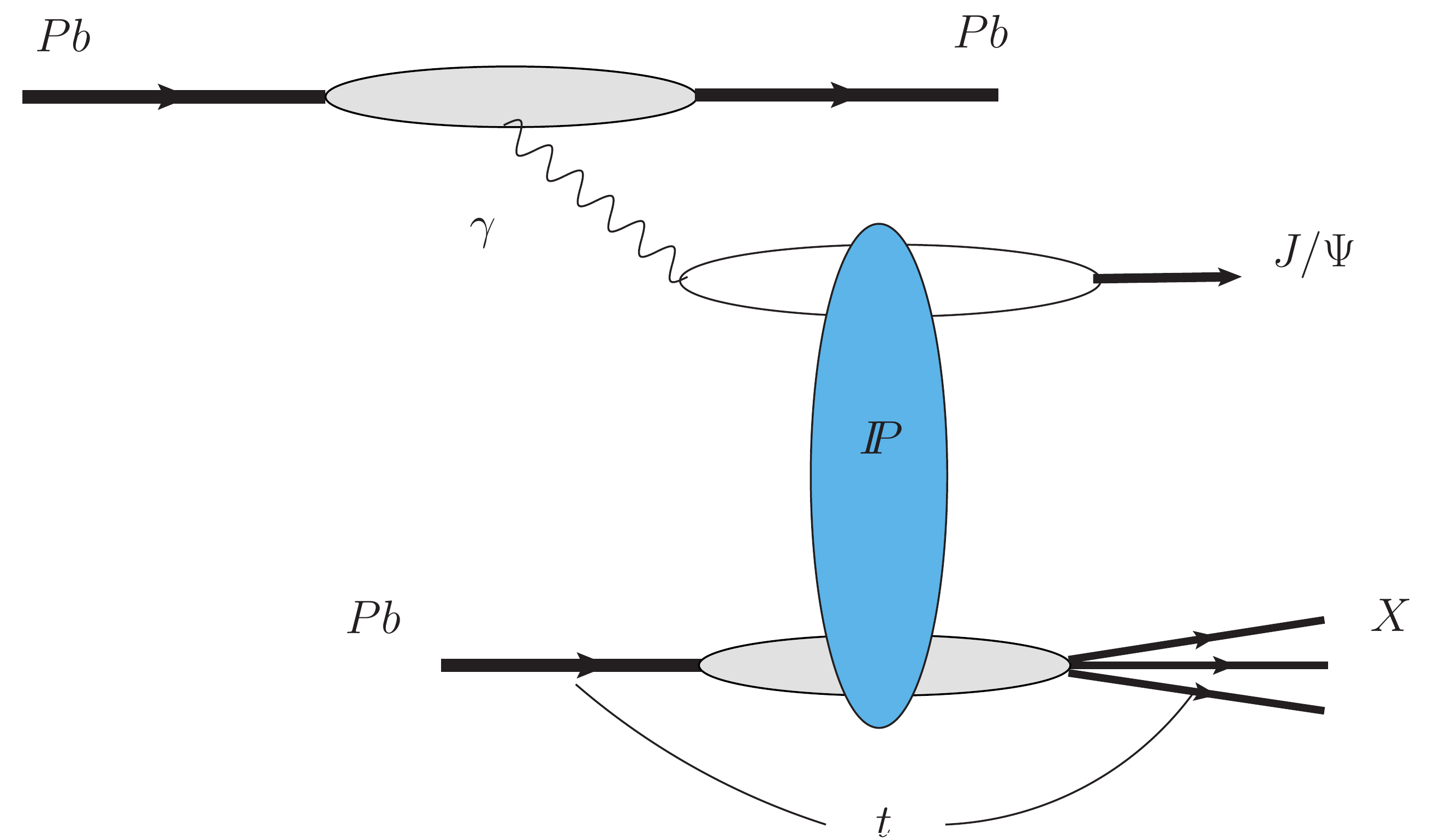}} \\ 
 (a) & (b)  
\end{tabular}                                                                                                                       
\caption{Typical diagrams for the (a) coherent and (b) incoherent $J/\Psi$ photoproduction in ultraperipheral $PbPb$ collisions.}
\label{fig:diagrama}
\end{figure}

Over the last decades, experiments at RHIC and LHC have collided a variety of nuclei over a wide range of energies, allowing to produce and characterize the properties of QGP as well as to study the production of different final states generated in ultraperipheral heavy ion collisions. In the coming years,  new LHC data at larger energies ($\sqrt{s} = 5.5$ and 10.6 TeV) and the implementation of the nuclear programme at the FCC ($\sqrt{s} = 39$ TeV) are expected to    
advance in our understanding of the nature of the hot and dense QCD matter produced in these collisions \cite{Dainese:2016gch}. However, the accuracy with which the properties of the QGP can be constrained in these future collisions strongly depends on the knowledge of the incoming nuclear wave functions at small - $x$.
In this paper, motivated by the studies performed in the Refs. \cite{vicmag,Lappi,armestoamir,Diego1,run2,heikkeplb,guzey_tdist,contreras,Luszczak:2017dwf,Guzey:2018tlk,Diego2,Sambasivam:2019gdd}	 for smaller center - of - mass energies, we will investigate the possibility
of determine the presence of gluon saturation effects and estimate the   magnitude of the associated non-linear corrections for the QCD dynamics in ultraperipheral $PbPb$ collisions for the energies of the next run of the  LHC ($\sqrt{s} = 5.5$ TeV) \cite{hl_lhc}, as well as for the energies of the  High -- Energy LHC ($\sqrt{s} = 10.6$ TeV) \cite{he_lhc} and Future Circular Collider ($\sqrt{s} = 39$ TeV) \cite{fcc}. In particular, we 
will consider the exclusive  photoproduction of $J/\Psi$ on heavy nuclei,  which is driven by the gluon content of the  nucleus and is strongly sensitive to non-linear effects (parton saturation). We 
will estimate the contribution of the coherent and incoherent  $J/\Psi$ processes, 
which provide different insights about the nuclear structure and the QCD dynamics at high energies \cite{Toll,Mantysaari:2016ykx,Mantysaari:2016jaz,cepila}. Such processes  are represented in Fig. \ref{fig:diagrama}, where the Pomeron ($\pom$) represents a color singlet exchange between the dipole and the target. If the nucleus scatters elastically,  the process is called coherent production, and the associated cross section measures the average spatial distribution of gluons in the target. On the other hand,  if the nucleus scatters inelastically, the process is denoted incoherent production.
 In this case, one sums over all final states of the target nucleus,
except those that contain particle production. The associated cross section probes the fluctuations and correlations in the nuclear gluon density.
In both cases, the final state is characterized by two rapidity gaps. As demonstrated in Refs. \cite{Toll,Mantysaari:2016ykx,Mantysaari:2016jaz,cepila}, the coherent production 
probes the averaged density profile of the gluon density, while the incoherent cross sections constrain the event - by - event fluctuations of the gluonic fields in the target.  In our analysis, 
  we will describe the nuclear profile taking into account of the possible states of nucleon configurations in the nuclear wave function, assuming  that each nucleon in the nucleus  has a Gaussian profile of width $B_p$, centered at random positions sampled from a Woods-Saxon nuclear profile \cite{Toll,cepila}. The numerical calculations will be performed using the Sar{\it t}re event generator proposed in Ref. \cite{Toll} and detailed in Ref. \cite{sartre}. In order to estimate the impact of the non - linear (saturation)  effects, we will compare the full predictions with those obtained disregarding these effects. As we will demonstrate below,  the total cross sections for coherent and incoherent processes, as well as the corresponding rapidity and transverse momentum distributions, are sensitive to the non  -  linear effects. 
Our results indicate that the study of the exclusive $J/\Psi$ photoproduction in ultraperipheral $PbPb$ collisions    at the LHC, HE - LHC and FCC can be useful  to discriminate between the saturation and non - saturation scenarios.

This paper is organized as follows. In the next Section, we present a brief review of the formalism used to estimate the coherent and incoherent cross sections as well the model  for the nuclear profile used in our calculations. In Section \ref{sec:results} we present our results for the coherent and incoherent cross sections, considering the kinematical range that will be probed by the LHC, HE - LHC and FCC. Finally, in Section \ref{sec:conc} we summarize our main conclusions.

\section{Formalism}
\label{form}

The coherent and incoherent $J/\Psi$ photoproduction in ultraperipheral $PbPb$ collisions are  represented by the diagrams shown in Fig. \ref{fig:diagrama}. As pointed before, the final state will be characterized by two rapidity gaps, i.e. the outgoing particles  are separated by a large region in rapidity in which there is no additional hadronic activity observed in the detector. In the case of coherent interactions (left panel), the nucleus scatters elastically and remains intact in the final state. In contrast, in incoherent interactions (right panel), the nucleus scatters inelastically, i.e., breaks up due to the $p_T$ ($=\sqrt{-t}$) kick given to the nucleus. Theoretically, it is expected that the coherent production dominates at small squared transverse momentum transfer $t$ ($|t|\cdot R^2/3 \ll 1$, where $R$ is the nuclear radius), with its signature being a sharp
forward diffraction peak. On the other hand, incoherent production is expected to dominate at large $t$ ($|t|\cdot R^2/3 \gg 1$), with the associated $t$-dependence being to a good accuracy the same as in the production off free nucleons. 
In ultraperipheral collisions, the $PbPb$ cross sections for the coherent and incoherent processes  can be written 
in a factorized form, 
given by the so called equivalent photon approximation \cite{epa}, with the differential cross sections being expressed as follows
\begin{eqnarray}
\frac{d\sigma_{coh}}{dy\,dt}  =  n_{Pb}(y) \, \cdot \, \left.\frac{d\sigma}{dt}(\gamma Pb 
\rightarrow J/\Psi  Pb;y)\right|_{coh} + n_{Pb}(-y) \, \cdot \, \left.\frac{d\sigma}{dt}(\gamma Pb 
\rightarrow J/\Psi  Pb; -y)\right|_{coh}\,\,\,,
\label{dsigdy_coh}
\end{eqnarray}
and
\begin{eqnarray}
\frac{d\sigma_{inc}}{dy\,dt}  =  n_{Pb}(y) \, \cdot \, \left.\frac{d\sigma}{dt}(\gamma Pb 
\rightarrow J/\Psi  X;y)\right|_{inc} + n_{Pb}(-y) \, \cdot \, \left.\frac{d\sigma}{dt}(\gamma Pb 
\rightarrow J/\Psi  X; -y)\right|_{inc}\,\,\,,
\label{dsigdy_inc}
\end{eqnarray}
where  $y$ is the rapidity of the $J/\Psi$ in the final state, which determines the photon energy $\omega$ in the collider frame and, consequently, the 
photon - nucleus center of mass energy  
$W = \sqrt{4 \omega E}$, where $E = \sqrt{s}/2$ and $\sqrt{s_{NN}}$ is the total collision energy per nucleon pair in the center-of-mass frame.
 As both incident nuclei 
act as a source of 
 photons \cite{upc},   the contributions associated to photon - Pomeron and Pomeron - photon interactions are taken into account in the above equations. Moreover, $n_{A}$ denotes the  equivalent photon 
spectrum  of the relativistic incident nucleus. As in our previous studies \cite{run2,Diego1,Diego2} we will assume   a point -- like form factor for the nucleus, which implies that \cite{upc}
 \begin{eqnarray}
   n_{A}(\omega) = \frac{2Z^{2}\alpha_{em}}{\pi }  
\left[
\xi K_{0}(\xi) K_{1}(\xi) -\frac{\xi^{2}}{2} \left( K_{1}^{2}(\xi) - K_{0}^{2}(\xi)  
\right )  \right]  ,
  \end{eqnarray}
where $
 \xi = \omega \left( 2R \right) / \gamma_{L}$, 
with $\gamma_L$ being the Lorentz factor.
In our analysis we are assuming that  the photons  emitted  are coherently radiated by the whole nucleus. Such condition imposes that the minimum photon wavelength must be greater than the nuclear radius. As a consequence, the photon virtuality must satisfy $Q^2 = -q^2 \le 1/R^2$, with the photon four -- momentum being $q^{\mu} = (\omega, \vec{q_{\perp}},q_z = \omega/v)$, where $\vec{q_{\perp}}$ is the transverse momentum of the photon in a given frame, where the projectile moves with velocity  $v$. It implies that $Q^2 = \omega^2/\gamma_L^2 + q_{\perp}^2$. The coherence condition limits the maximum energy of the photon to $\omega < \omega_{\mbox{max}}  \approx \gamma_L/R$ and the perpendicular component of its momentum to ${q_{\perp}} \le 1/R$. Therefore, the coherence condition  sets an upper limit on the transverse momentum of the photon emitted by the nucleus, which should satisfy $q_{\perp} \le 1/R$, being $\approx 28$ MeV for $Pb$ beams. Consequently, the photon virtuality can be neglected and the photons can be considered as being real. The maximum photon energy can also be derived considering that the maximum possible momentum in the longitudinal direction is modified by the Lorentz factor, $\gamma_L$, due to the Lorentz contraction of the nucleus in that direction. It implies $\omega_{\mbox{max}} \approx \gamma_L/R$ and, consequently, $W^{\mbox{max}} = \sqrt{2\,\omega_{\mbox{max}}\, \sqrt{s_{NN}}}$. Considering the values of $\sqrt{s_{NN}}$  for $PbPb$ collisions at the LHC ($\sqrt{s_{NN}} = 5.5$ TeV)
  and FCC ($\sqrt{s_{NN}} = 39$ TeV), we obtain that
 the maximum photon -- nucleon center -- of -- mass energy, $W^{\mbox{max}}$,  reached in these collisions are $0.95$ TeV and  $6.8$ TeV, respectively.
Such values are much larger than those studied at HERA and that will be accessed in the future electron -- ion collider. Therefore, the study of photon -- nucleus interactions at LHC and FCC will allow us to probe the QCD dynamics in a unexplored kinematical range. As pointed out in the Introduction, to establish the dynamics at small - $x$ is fundamental to the success of the heavy ion physics program.

The main input in Eqs. (\ref{dsigdy_coh}) and (\ref{dsigdy_inc}) are the differential cross sections, $d\sigma/dt$, for the coherent and  incoherent interactions. In order to estimate these quantities we will take into account the distinct nucleon configurations of the nucleus and average over all possible configurations. 
For coherent interactions, in which the nucleus is required to remain in its ground state, the average over the configurations of the nuclear wave function, denoted by $\left\langle ... \right\rangle$ hereafter, is taken at the level of the scattering amplitude. Consequently, the coherent cross section is obtained by averaging the amplitude before squaring it and the differential distribution will be given by
\begin{equation}\label{eq:xsec-coh}
  \left.\frac{d\sigma^{\gamma Pb \rightarrow J/\Psi \,Pb}}{dt}\right|_{coh} =
  \frac{1}{16\pi}\left| \left\langle \mathcal{A}(x, \Delta) \right\rangle \right|^2\,\,,
\end{equation}
where $x = (M^2  -t)/(W^2)$, with $M$ being the $J/\Psi$ mass, and $\Delta = \sqrt{-t}$ is the momentum transfer.
On the other hand, for incoherent interactions, the average over configurations is at the cross section level. In this case, the nucleus can break up and the resulting incoherent cross section will be proportional to the variance of the amplitude with respect to the nucleon configurations of the nucleus, i.e., it will measure the fluctuations of the gluon density inside the nucleus. The differential cross sections for incoherent interactions will be expressed as follows:
\begin{equation}\label{eq:xsec-inc}
  \left.\frac{d\sigma^{\gamma Pb \rightarrow J/\Psi\,X}}{dt}\right|_{inc} = \frac{1}{16\pi}
  \left(  \left\langle\left|  \mathcal{A}(x, \Delta)  \right|^2 \right\rangle  - \left| \left\langle \mathcal{A}(x, \Delta) \right\rangle \right|^2\right),
\end{equation}
where $X = Pb^*$ represents the dissociative state. In our calculations we will include the skewedness correction by multiplicating the coherent and incoherent cross sections by the factor $R_g^2$ as given in Ref. \cite{Shuvaev:1999ce}.
In the color dipole formalism,  the  scattering amplitude $\mathcal{A}(x, \Delta)$ can be factorized in terms of the fluctuation of the  photon into a $q \bar{q}$ color dipole, the dipole-nucleus scattering by a color singlet exchange  and the recombination into the exclusive final state $J/\Psi$, being given by
\begin{eqnarray}
 {\cal A}({x},\Delta)  =  i
\,\int d^2\rr \int \frac{dz}{4\pi} \int \, d^2\rb \, e^{-i[\rb -(1-z)\rr].\rd}
 \,\, (\Psi^{V*}\Psi)  \,\,\frac{d\sigma_{dA}}{d^2\rb}({x},\rr,\rb)
\label{amp}
\end{eqnarray}
where  $(\Psi^{V*}\Psi)$ denotes the wave function overlap between the  photon and the $J/\Psi$ wave functions, which will be described using the Boosted Gaussian model (For details see e.g. Ref. \cite{run2}). The variables  $\rr$ and $z$ are the dipole transverse radius and the momentum fraction of the photon carried by a quark (an antiquark carries then $1-z$), respectively, and   $\rb$ is the impact parameter of the dipole relative to the target. Moreover,  ${d\sigma_{dA}}/{d^2\rb}$  is the dipole-nucleus cross section (for a dipole at  impact parameter $\rb$) which encodes all the information about the hadronic scattering, and thus about the non-linear and quantum effects in the hadron wave function.
 How to treat the dipole - nucleus interaction  is still an open question 
due to the complexity of  the impact parameter dependence. In principle, ${d\sigma_{dA}}/{d^2\rb}$ can be derived using the
Color Glass Condensate (CGC) formalism \cite{CGC}, which is characterized by the infinite hierarchy of equations, the so called Balitsky-JIMWLK equations \cite{BAL,CGC}, which reduces in the mean field approximation to the Balitsky-Kovchegov (BK) equation \cite{BAL,kov}.

In our analysis, following the studies presented in Refs.  \cite{run2,Toll,Diego1,cepila},  we will describe the dipole - nucleus cross section using the
Glauber-Gribov formalism \cite{glauber,gribov,mueller}, which implies that ${d\sigma_{dA}}/{d^2\rb}$ is given by
\begin{eqnarray}
\frac{d\sigma_{dA}}{d^2\rb} = 2\,\left( 1 - \exp \left[-\frac{1}{2}  \, \sigma_{dp}(x,\rr^2) \,T_A(\rb)\right]\right) \,\,,
\label{enenuc}
\end{eqnarray}
where $\sigma_{dp}$ is the dipole-proton cross section and $T_A(\rb)$ is  the nuclear profile function. 
We will describe the nuclear profile $T_A(\rb)$ taking into account of all possible states of nucleon configurations in the nuclear wave function. Following Refs. \cite{Toll,cepila}, we will assume that each nucleon in the nucleus  has a Gaussian profile of width $B_p$, centered at random positions $\rb_i$ sampled from a Woods-Saxon nuclear profile as follows
\begin{equation}\label{eq:Ths0}
  T_A(\rb) = \frac{1}{2\pi B_p} \sum_{i=1}^{A} \exp\left[ - \frac{(\rb - \rb_i)^2}{2B_p} \right] \,\,.
\end{equation}
Moreover, as in Ref. \cite{Toll}, the dipole - proton cross section  will be given by
\begin{equation}
\sigma_{dp}(x,\rr^2) = \frac{\pi^2 r^{2}}{ N_{c}} \alpha_{s}(\mu^{2}) \,\,xg\left(x, \mu^2 = \frac{C}{r^{2}} + 
\mu_{0}^{2}\right) \,\,\,
\end{equation}
where the gluon distribution evolves via DGLAP equation, with the initial condition at $\mu_{0}^{2}$  taken to be $
xg(x,\mu_{0}^{2}) =  A_{g}x^{-\lambda_{g}} (1-x)^{6}$.  
In this work, we 
assume the parameters $B_p, A_g, \lambda_g, C$ and $\mu_0^2$ obtained in Ref. \cite{ipsat_heikke} for the IP-SAT model.  
We will denote by b - Sat the predictions derived using  Eq. (\ref{enenuc}) as input in the calculations.
In order to estimate the impact of non-linear corrections to the QCD dynamics, we also will estimate the observables assuming that the dipole - nucleus cross section is given by:
\begin{eqnarray}
\frac{d\sigma_{dA}}{d^2\rb} =  \sigma_{dp}(x,\rr^2) \,T_A(\rb) \,\,,
\label{enenuc_lin}
\end{eqnarray}
which disregards the effect of the multiple elastic dipole rescatterings. The associated predictions will be denoted by b - Non Sat hereafter. For this case, 	 we 
assume the parameters $B_p, A_g, \lambda_g, C$ and $\mu_0^2$ obtained in Ref. \cite{ipsat_heikke} for the IP-NONSAT model.

 \begin{figure}[t]
\begin{tabular}{ccc}

 {\includegraphics[width=0.33\textwidth]{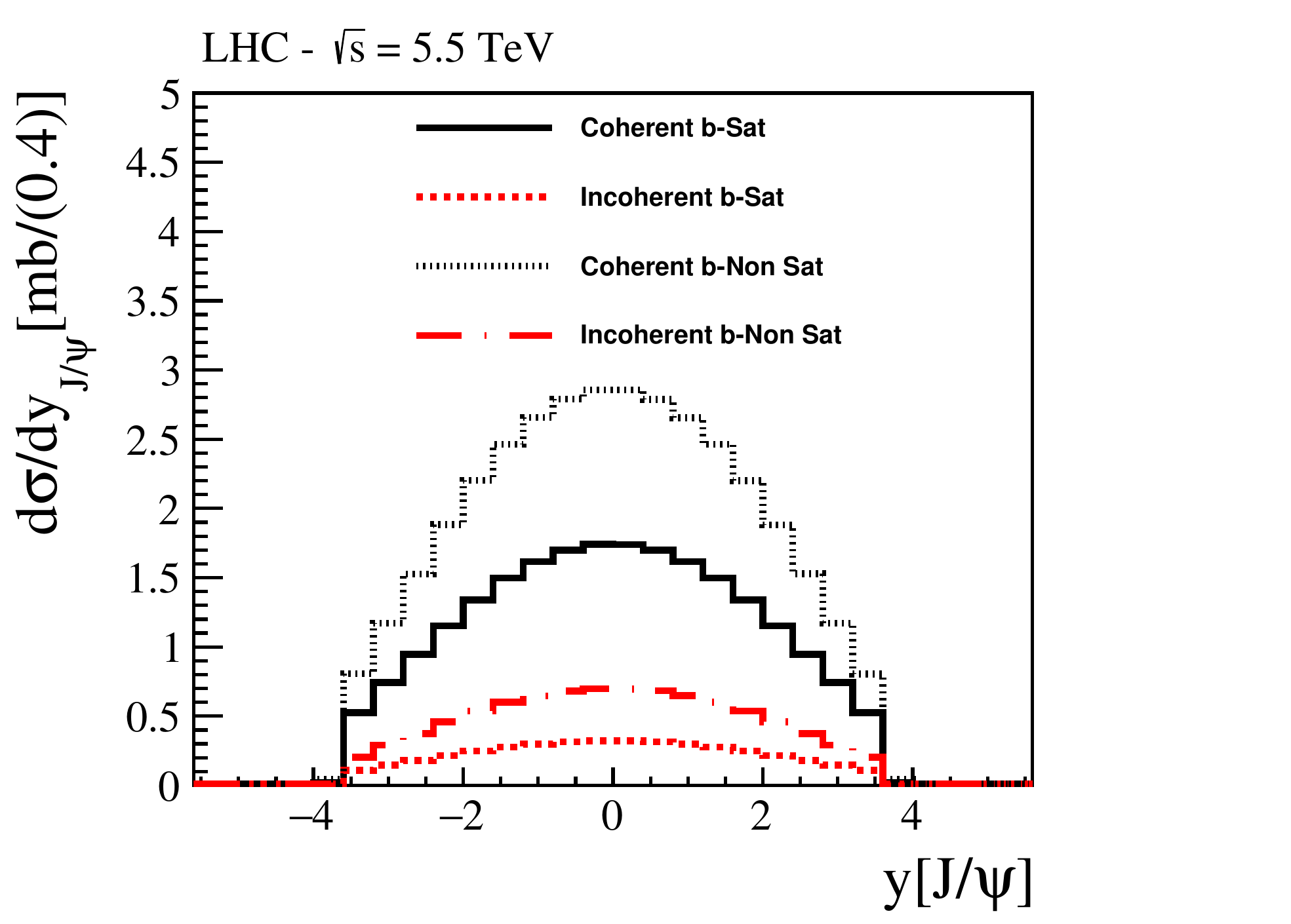}} & 
{\includegraphics[width=0.33\textwidth]{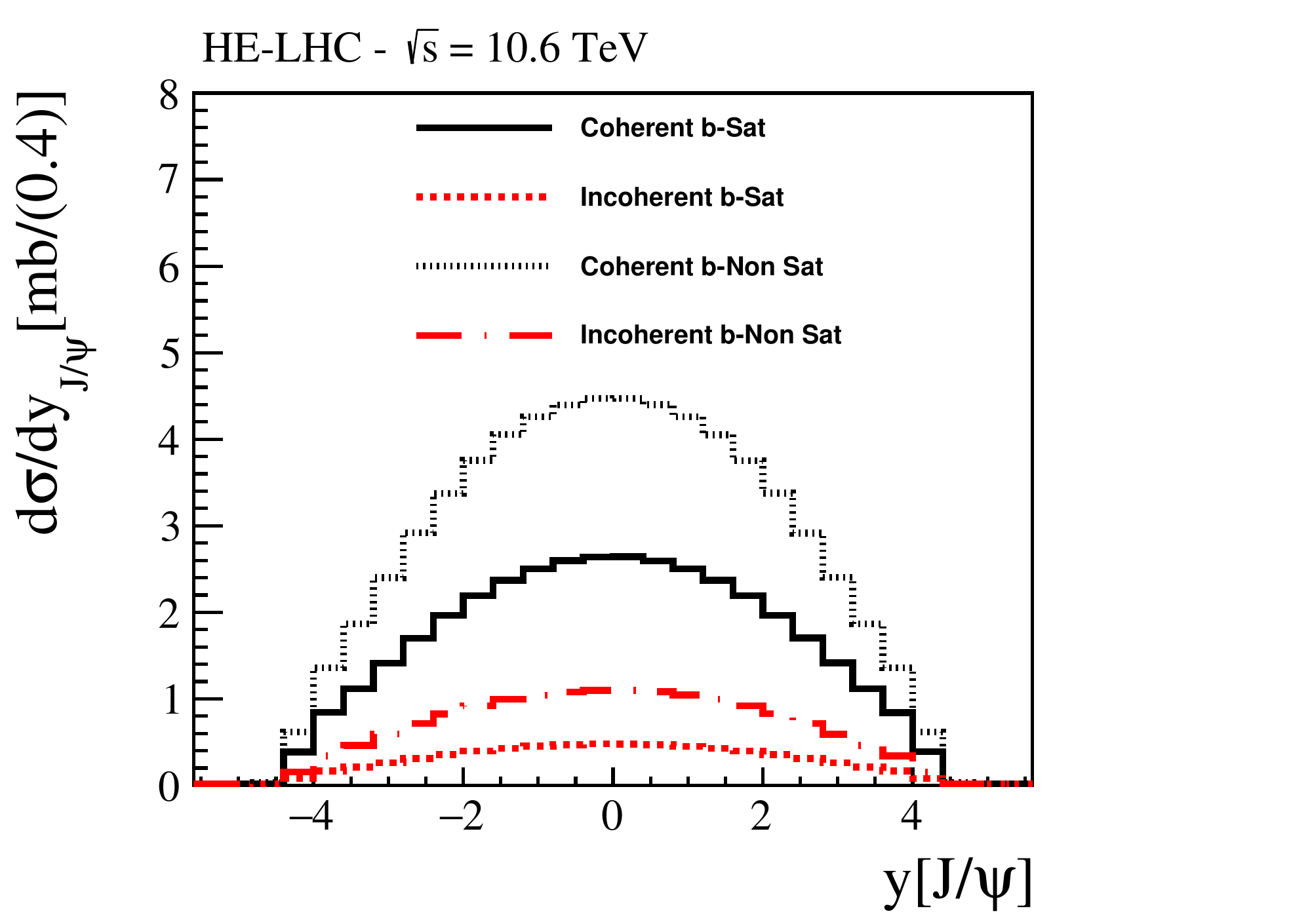}} & {\includegraphics[width=0.33\textwidth]{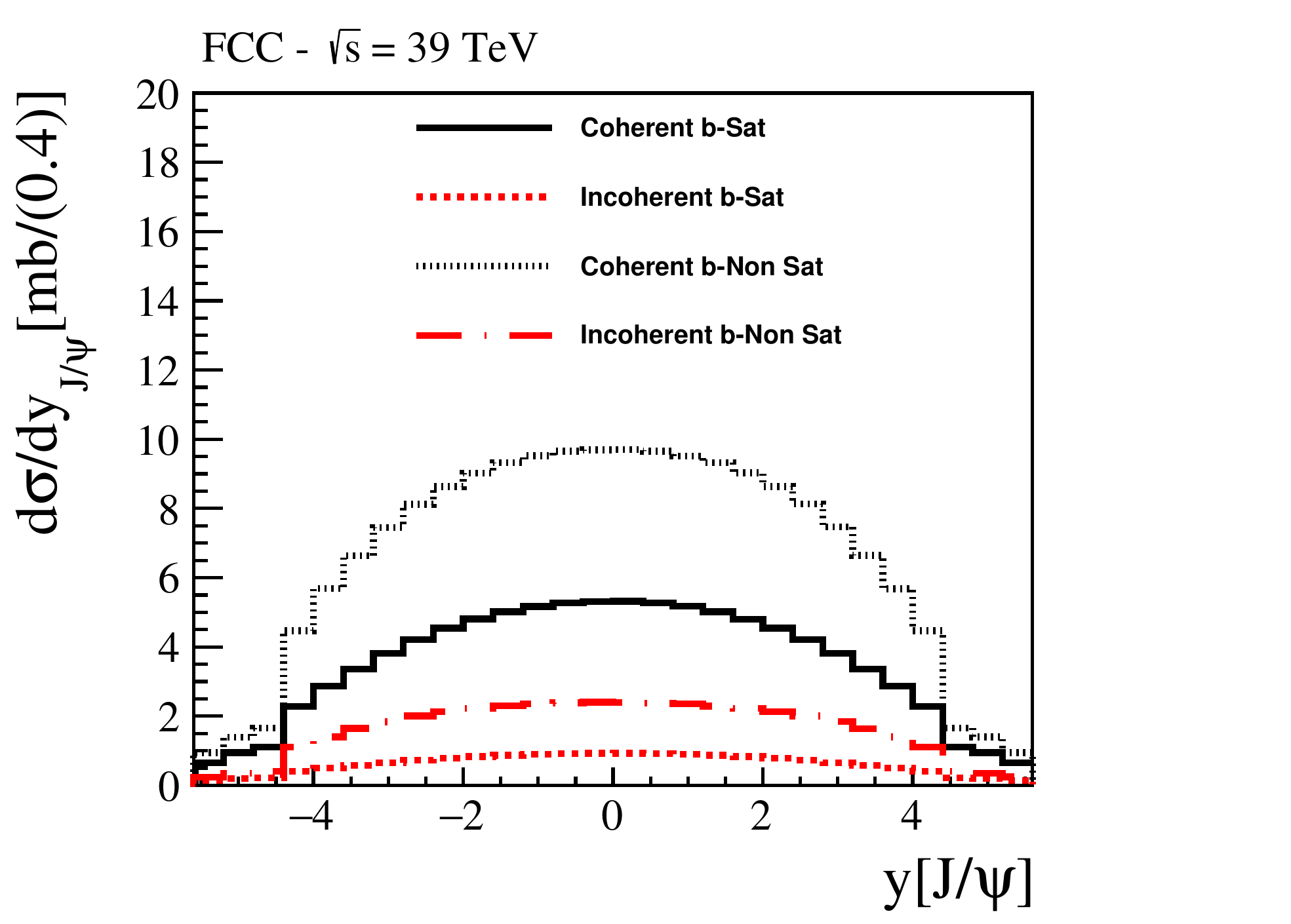}}    \\ 
 (a) & (b) & (c) 
\end{tabular}                                                                                                                       
\caption{Rapidity distributions for the coherent and incoherent $J/\Psi$ photoproduction in $PbPb$ collisions for the (a) LHC, (b) HE - LHC and (c) FCC energies.}
\label{fig:rapidity}
\end{figure}

\section{Results}
\label{sec:results}

In what follows, we will present  predictions for the coherent and incoherent $J/\Psi$ photoproduction in $PbPb$ collisions for the energies of the next run of the LHC ($\sqrt{s} = 5.5$ TeV) \cite{hl_lhc}, as well as for the energies of the  High -- Energy LHC ($\sqrt{s} = 10.6$ TeV) \cite{he_lhc} and Future Circular Collider ($\sqrt{s} = 39$ TeV) \cite{fcc}. The numerical calculations will be performed using the Sar{\it t}re event generator \cite{sartre}. In order to perform the averages present in the coherent and incoherent cross sections, we have considered 500 distinct nucleon configurations. As demonstrated in \cite{Toll}, this number of configurations is enough to obtain a good description of the cross sections for $|t| \le 0.08$ GeV$^2$, which is the range of interest in our study.

Initially, let's estimate the rapidity distribution, which is one of the main observables that 
can be directly measured at the LHC and FCC. The predictions for coherent and incoherent interactions can be obtained from Eqs. (\ref{dsigdy_coh}) and (\ref{dsigdy_inc}) by integrating over all values of  $t$. The results are presented in Fig. \ref{fig:rapidity}. One has that the coherent interactions dominate, in agreement with the results presented in Refs. \cite{heikkeplb,Diego1,contreras} for smaller center - of - mass energies. Such result is expect, since the coherent production is characterized by a sharp
forward diffraction peak, being much larger than the incoherent one for small values of $|t|$ (see below). 	Moreover, we have that the values of the rapidity distribution for midrapidity increase with the energy, with the increasing being dependent on the modeling of the QCD dynamics. We have that the b-Sat predictions are a factor $\gtrsim 1.5$ smaller than the b-Non Sat one. The associated cross sections are presented in Table \ref{tab:cross} by integrating over the full rapidity range as well as over the typical ranges covered by  central and forward detectors.
The cross sections are of the order of mb, which implies that the number of events per year at the LHC / HE-LHC / FCC will be larger than $10^6/\,10^7/\,10^8$, if we assume the expected integrated luminosity as being  ${\cal{L}} = 3.0 /\,10 /\,110$ $nb^{-1}$ \cite{hl_lhc,he_lhc,fcc}.  Such large number of events implies that a detailed analysis of the coherent and incoherent processes is, in principle, feasible. Our results indicate that the measurement of the rapidity distribution can be useful to discriminate between the b-Sat and b-Non Sat scenarios.

\begin{table}[t]
\centering
\begin{tabular}{|c||c|c||c|c||c|c|}\hline 
{\bf PbPb Collisions} &   \multicolumn{2}{|c||}{\bf $\sqrt{s}=5.5$ TeV} &    \multicolumn{2}{|c||}{\bf $\sqrt{s} =  10.6$ TeV} &     \multicolumn{2}{|c|}{\bf $\sqrt{s} = 39$ TeV}      \\ \hline         
{\bf Dipole Model}  &  b-Sat & b-Non Sat  &  b-Sat & b-Non Sat  &  b-Sat & b-NonSat  \\ \hline        
\hline
{\bf Coherent (Total)}  &  22.6 & 36.8     &  39.5  & 67.0      &  98.7  & 184.4     \\ \hline
{\bf Coherent ($|y|<2.0$)}  &  15.8 & 25.9     &  24.6  & 41.9       &  51.1  & 94.4      \\ \hline 
{\bf Coherent ($2.0 < y < 4.5$)}  &  3.4 & 5.4     &  7.4  & 12.6       &  21.4  & 41.5      \\ \hline 
\hline
{\bf Incoherent (Total)}  &  4.2 & 9.0     &  7.2  & 16.5       &  17.3  & 45.6      \\ \hline          
                         {\bf Incoherent ($|y|<2.0$)}  &  2.9 & 6.3     &  4.4  & 10.3       &  8.9  &  23.3      \\ \hline 
{\bf Incoherent ($2.0<y<4.5$)}  &  0.7 & 1.3     &  1.4  & 3.1       &  3.7  &  10.2      \\ \hline 
\hline  
\end{tabular} 
\caption{Cross sections, in mb, for the coherent and incoherent $J/\Psi$ photoproduction in $PbPb$ collisions at the LHC, HE - LHC and FCC considering the b-Sat and b-Non Sat dipole models.}
\label{tab:cross}
\end{table}

\begin{figure}[t]
\begin{tabular}{ccc}
{\includegraphics[width=0.33\textwidth]{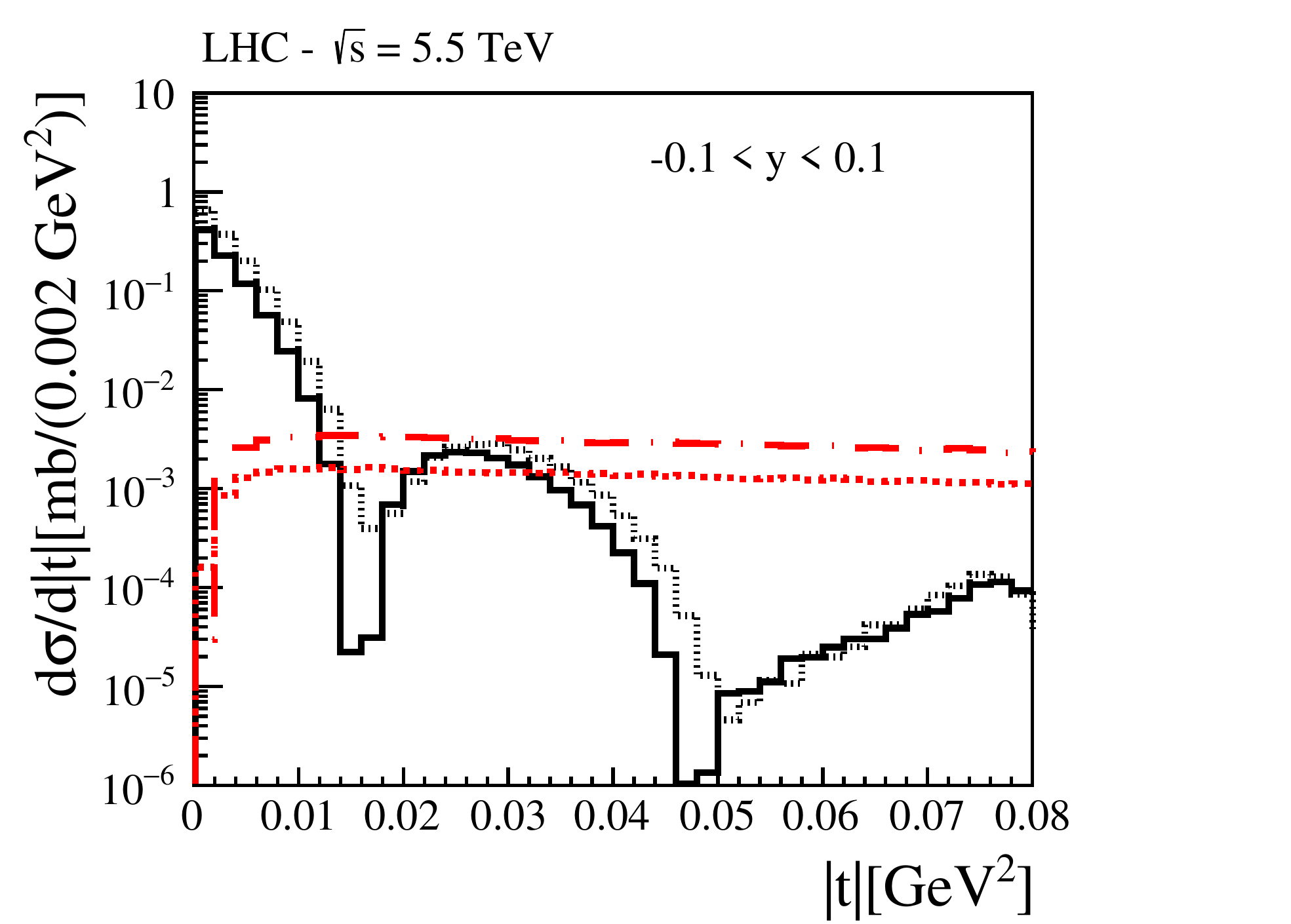}} &  {\includegraphics[width=0.33\textwidth]{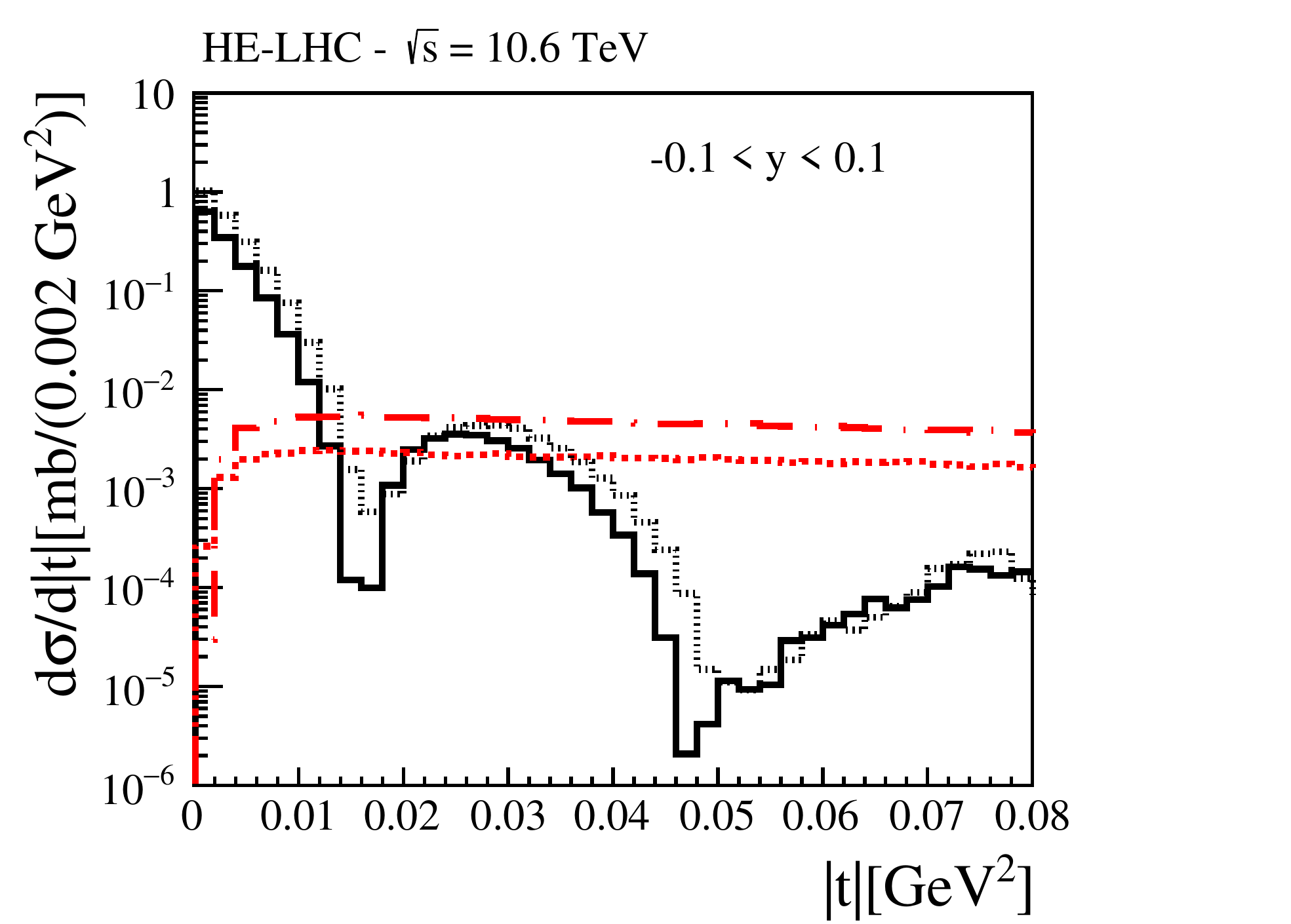}} & 
 {\includegraphics[width=0.33\textwidth]{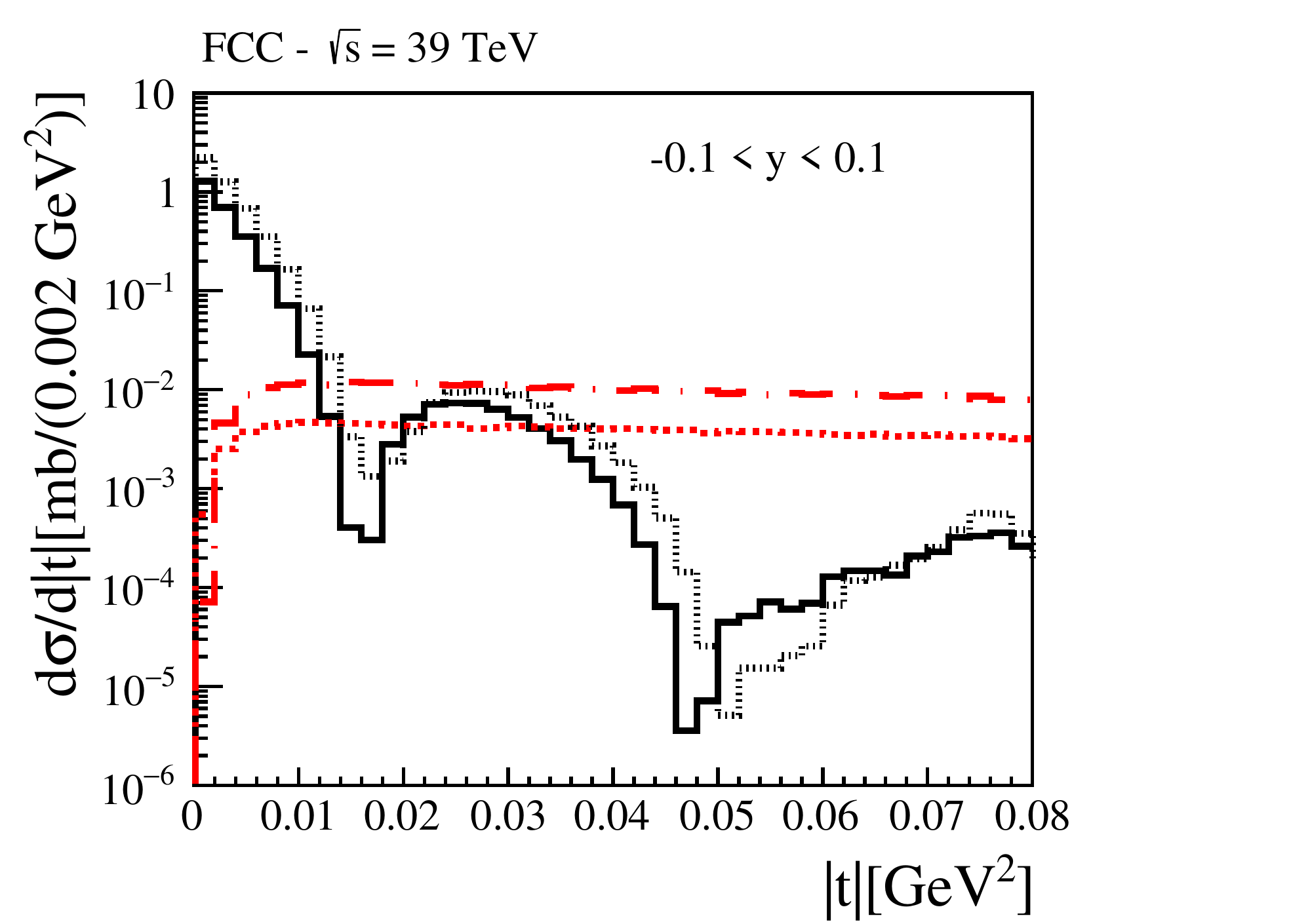}}  \\
{\includegraphics[width=0.33\textwidth]{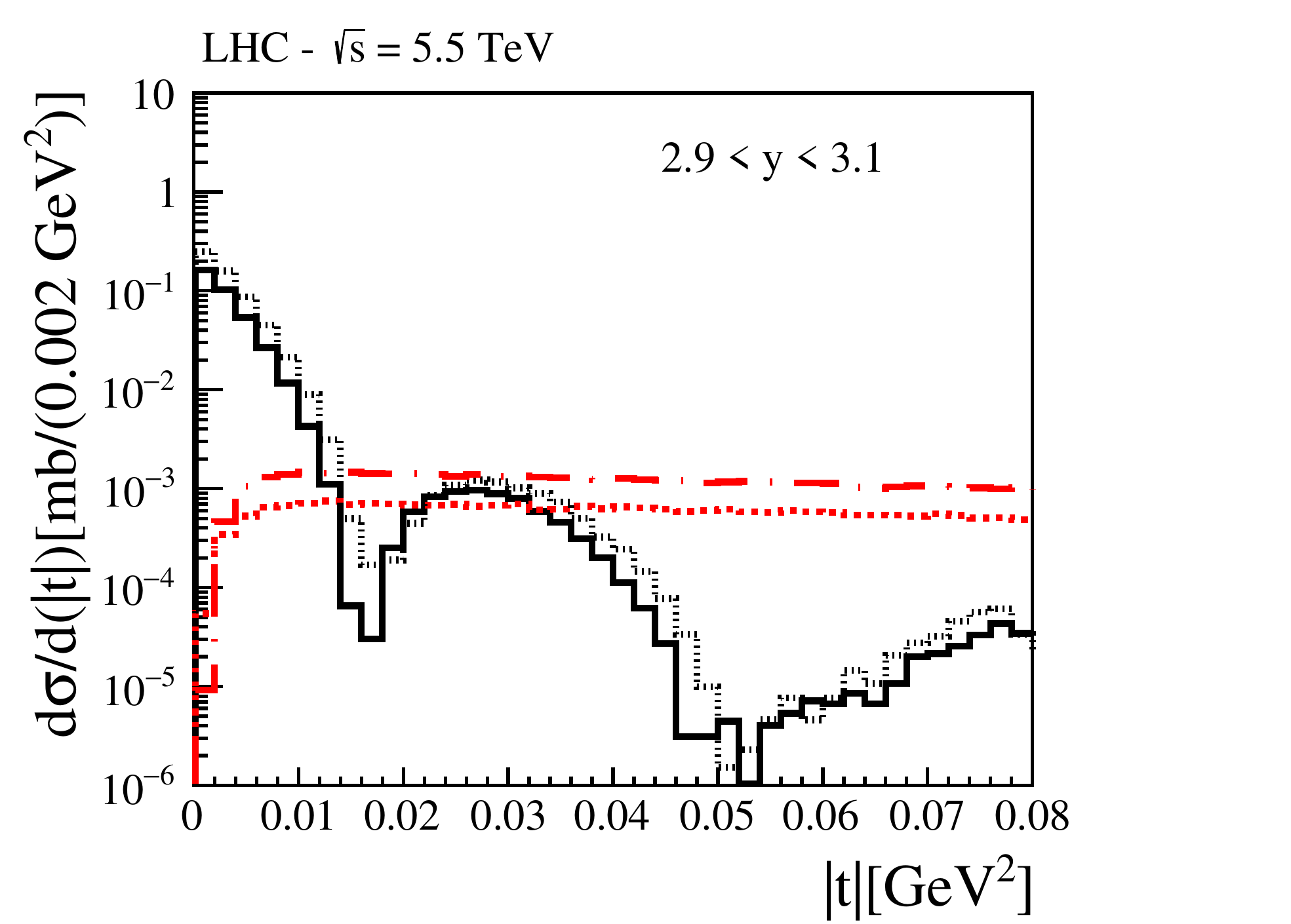}} &  {\includegraphics[width=0.33\textwidth]{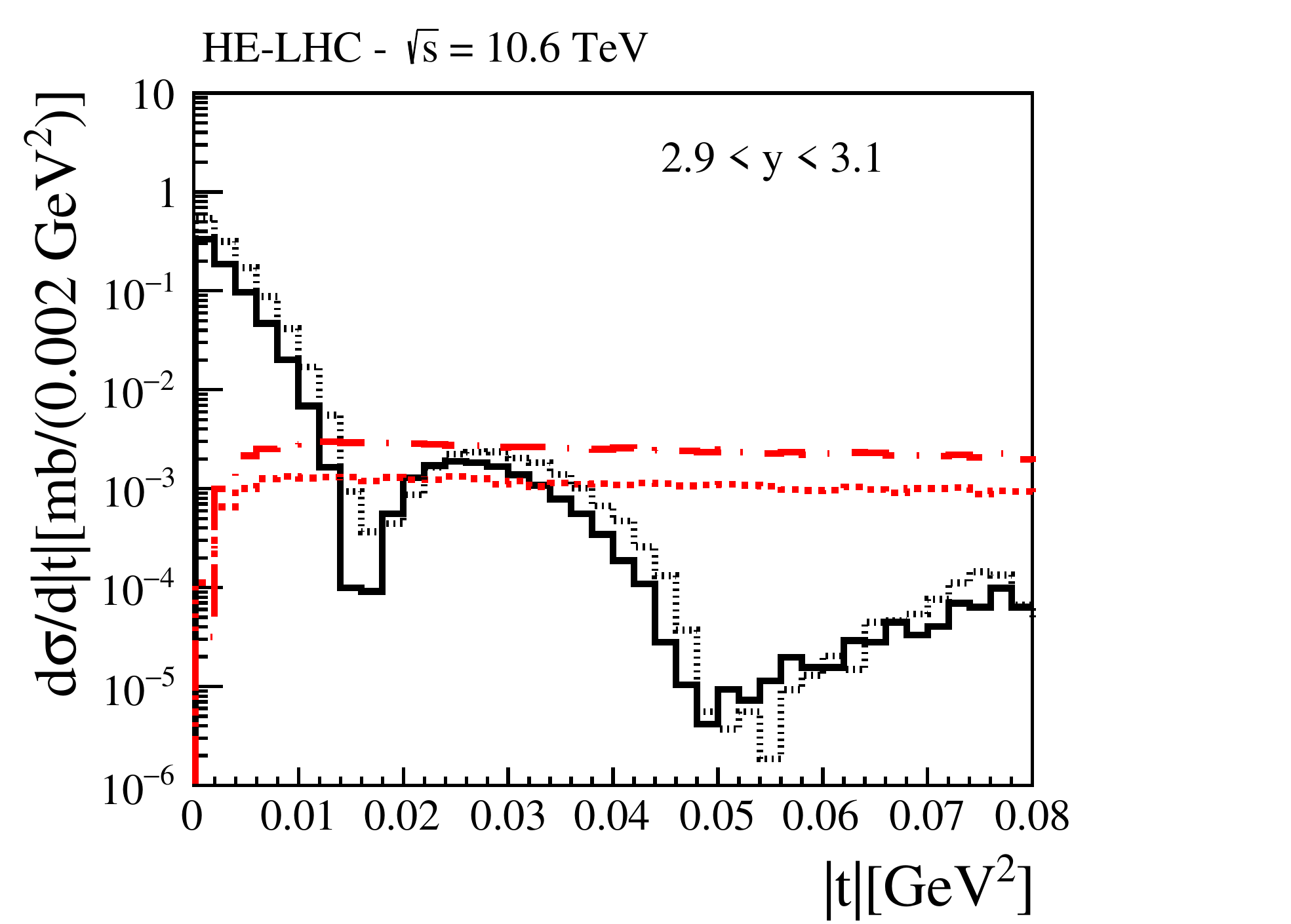}} & 
 {\includegraphics[width=0.33\textwidth]{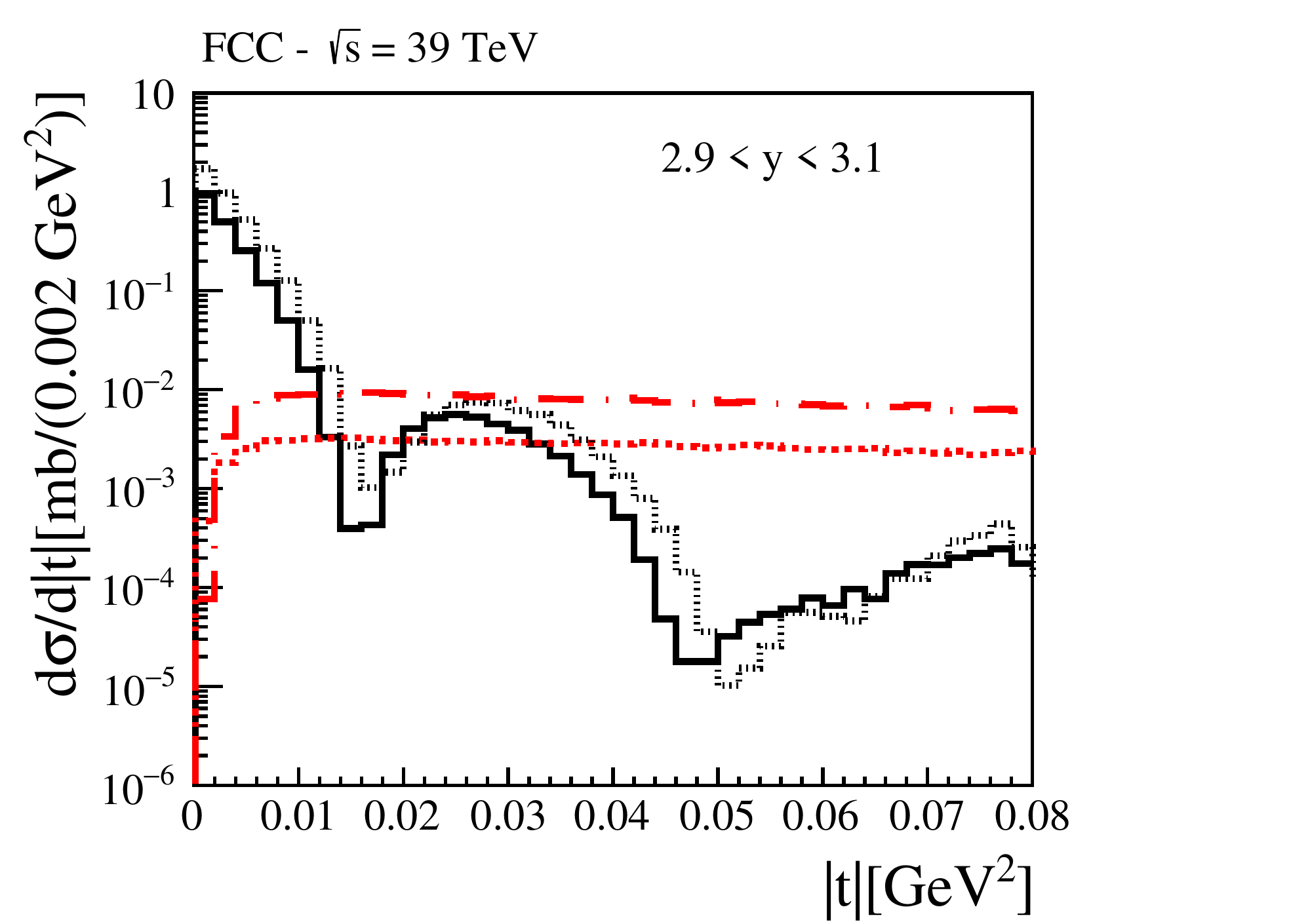}}  \\
\end{tabular}                                                                                                                       
\caption{Transverse momentum distributions for the coherent and incoherent $J/\Psi$ photoproduction in $PbPb$ collisions for the  LHC, HE - LHC and FCC energies   considering the central (upper panels) and forward (lower panels) rapidity ranges.}
\label{fig:tdist}
\end{figure}

Another observable of interest is the squared  momentum transfer ($t$) distribution for a fixed rapidity. As demonstrated in previous studies \cite{armestoamir,Diego1,Diego2}, such distribution is an important alternative to probe the QCD dynamics at high energies and provide  information about the spatial distribution of the gluons in the target and about fluctuations  of the  color fields. Our predictions are presented in Fig. \ref{fig:tdist}
for distinct energies considering central (upper panels) and forward (lower panels) rapidities. As expected from previous studies \cite{Diego1,heikkeplb}, the coherent dominates at small - $|t|$ and the incoherent ones at large values of the momentum transfer, which is associated to the fact that increasing of the momentum kick given to the nucleus the probability that  it  breaks up becomes larger. As a consequence, the $J/\Psi$ production at large - $|t|$ is dominated by incoherent processes. In addition, the  coherent cross sections clearly exhibit the typical diffractive pattern  and is characterized by a sharp forward diffraction peak. In contrast, the incoherent cross section is characterized by a flat $t$ - dependence, decreasing when $|t| \rightarrow 0$. 
Regarding the impact of the saturation effects, one has that the normalization of the incoherent predictions is modified by the non - linear effects, with the difference between the b-Sat and b-Non Sat predictions increasing with the energy. A similar effect is also observed in the coherent case. However, for the coherent processes, the position of the dips is sensitive to the presence of the saturation effects, in agreement with the results obtained in Refs. \cite{Diego1,Diego2}. Our results indicate that the position of the second dip is dependent on description of the QCD dynamics, with the predictions becoming more distinct at larger energies. However, it is important to emphasize that the experimental separation of coherent processes at large - $|t|$ is still a challenge due to the dominance of the incoherent interactions. An alternative is the detection of the fragments of the nuclear breakup produced in the incoherent processes. e.g. the detection of emitted neutrons by zero - degree calorimeters \cite{Caldwell,Toll}.

\section{Summary}
\label{sec:conc}

Ultraperipheral heavy ion collisions at the LHC and FCC are an important alternative to constrain the QCD dynamics at high energies and, consequently, the description of the initial conditions for central and semi - central collisions. In particular, the increasing in the center - of - mass energy and integrated luminosity in the forthcoming experiments opens up new opportunities to probe the nuclear wave function in a unexplored energy range, where non - linear (saturation) effects are expected to significantly contribute. In this paper, we have performed a detailed investigation of the coherent and incoherent $J/\Psi$ photoproduction in $PbPb$ collisions considering  
the possible states of nucleon configurations in the nuclear wave function and taking into account of the non - linear corrections to the QCD dynamics. Moreover, a comparison with the results derived disregarding these corrections was also presented. We have derived predictions for the cross sections of the coherent and incoherent processes considering the rapidity ranges covered by central and forward detectors,  which demonstrated that the event rates of these processes are very large and that they are sensitive to saturation effects. Moreover, predictions for the rapidity and transverse momentum distributions were presented. In particular, these results indicate that the experimental analysis of the transverse momentum distribution is  useful to discriminate between different approaches for the QCD dynamics as well to improve our description of the gluon saturation effects. Finally, our results indicate  that a future experimental analysis of the coherent and incoherent processes will be useful to improve our understanding of the QCD dynamics at high energies.

\begin{acknowledgments}
VPG acknowledge useful discussions about coherent and incoherent interactions with Jan Cepila, Michal Krelina and Wolfgang Schafer.
This work was  partially financed by the Brazilian funding
agencies CNPq, CAPES,  FAPERGS, FAPERJ and INCT-FNA (processes number 
464898/2014-5 and 88887.461636/2019-00).
\end{acknowledgments}

\hspace{1.0cm}

\end{document}